\PassOptionsToPackage{unicode}{hyperref}
\PassOptionsToPackage{hyphens}{url}
\PassOptionsToPackage{dvipsnames,svgnames,x11names}{xcolor}
\documentclass[
]{article}

\usepackage{amsmath,amssymb}
\usepackage{iftex}
\ifPDFTeX
  \usepackage[T1]{fontenc}
  \usepackage[utf8]{inputenc}
  \usepackage{textcomp} % provide euro and other symbols
\else % if luatex or xetex
  \usepackage{unicode-math}
  \defaultfontfeatures{Scale=MatchLowercase}
  \defaultfontfeatures[\rmfamily]{Ligatures=TeX,Scale=1}
\fi
\usepackage{lmodern}
\ifPDFTeX\else  
\fi
\IfFileExists{upquote.sty}{\usepackage{upquote}}{}
\IfFileExists{microtype.sty}{% use microtype if available
  \usepackage[]{microtype}
  \UseMicrotypeSet[protrusion]{basicmath} % disable protrusion for tt fonts
}{}
\makeatletter
\@ifundefined{KOMAClassName}{% if non-KOMA class
  \IfFileExists{parskip.sty}{%
    \usepackage{parskip}
  }{% else
    \setlength{\parindent}{0pt}
    \setlength{\parskip}{6pt plus 2pt minus 1pt}}
}{% if KOMA class
  \KOMAoptions{parskip=half}}
\makeatother
\usepackage{xcolor}
\ifx\paragraph\undefined\else
  \let\oldparagraph\paragraph
  \renewcommand{\paragraph}[1]{\oldparagraph{#1}\mbox{}}
\fi
\ifx\subparagraph\undefined\else
  \let\oldsubparagraph\subparagraph
  \renewcommand{\subparagraph}[1]{\oldsubparagraph{#1}\mbox{}}
\fi

\usepackage{longtable,booktabs,array}
\usepackage{calc} % for calculating minipage widths
\usepackage{etoolbox}
\makeatletter
\patchcmd\longtable{\par}{\if@noskipsec\mbox{}\fi\par}{}{}
\makeatother
\IfFileExists{footnotehyper.sty}{\usepackage{footnotehyper}}{\usepackage{footnote}}
\makesavenoteenv{longtable}
\usepackage{graphicx}
\makeatletter
\def\maxwidth{\ifdim\Gin@nat@width>\linewidth\linewidth\else\Gin@nat@width\fi}
\def\maxheight{\ifdim\Gin@nat@height>\textheight\textheight\else\Gin@nat@height\fi}
\makeatother
\setkeys{Gin}{width=\maxwidth,height=\maxheight,keepaspectratio}
\makeatletter
\def\fps@figure{htbp}
\makeatother
\NewDocumentCommand\citeproctext{}{}

\makeatletter
 \let\@cite@ofmt\@firstofone
 \def\@biblabel#1{}
 \def\@cite#1#2{{#1\if@tempswa , #2\fi}}
\makeatother
\newlength{\cslhangindent}
\newlength{\csllabelwidth}
\newenvironment{CSLReferences}[2] % #1 hanging-indent, #2 entry-spacing
 {\begin{list}{}{%
  \setlength{\itemindent}{0pt}
  \setlength{\leftmargin}{0pt}
  \setlength{\parsep}{0pt}
  \ifodd #1
   \setlength{\leftmargin}{\cslhangindent}
   \setlength{\itemindent}{-1\cslhangindent}
  \fi
  \setlength{\itemsep}{#2\baselineskip}}}
 {\end{list}}
\usepackage{calc}

\usepackage{arxiv}
\usepackage{orcidlink}
\usepackage{amsmath}
\usepackage[T1]{fontenc}
\makeatletter
\@ifpackageloaded{caption}{}{\usepackage{caption}}
\AtBeginDocument{%
\ifdefined\contentsname
  \renewcommand*\contentsname{Table of contents}
\else
  \newcommand\contentsname{Table of contents}
\fi
\ifdefined\listfigurename
  \renewcommand*\listfigurename{List of Figures}
\else
  \newcommand\listfigurename{List of Figures}
\fi
\ifdefined\listtablename
  \renewcommand*\listtablename{List of Tables}
\else
  \newcommand\listtablename{List of Tables}
\fi
\ifdefined\figurename
  \renewcommand*\figurename{Figure}
\else
  \newcommand\figurename{Figure}
\fi
\ifdefined\tablename
  \renewcommand*\tablename{Table}
\else
  \newcommand\tablename{Table}
\fi
}
\@ifpackageloaded{float}{}{\usepackage{float}}
\floatstyle{ruled}
\@ifundefined{c@chapter}{\newfloat{codelisting}{h}{lop}}{\newfloat{codelisting}{h}{lop}[chapter]}
\floatname{codelisting}{Listing}

\makeatother
\makeatletter
\@ifpackageloaded{caption}{}{\usepackage{caption}}
\@ifpackageloaded{subcaption}{}{\usepackage{subcaption}}
\makeatother
\ifLuaTeX
  \usepackage{selnolig}  % disable illegal ligatures
\fi
\usepackage{bookmark}

\IfFileExists{xurl.sty}{\usepackage{xurl}}{} % add URL line breaks if available
\hypersetup{
  pdftitle={Forecasting Densities of Fatalities from State-based Conflicts using Observed Markov Models},
  pdfauthor={David Randahl; Johan Vegelius},
  pdfkeywords={Markov models, Forecasting, Density
forecasts, Fatalities},
  colorlinks=true,
  linkcolor={blue},
  filecolor={Maroon},
  citecolor={Blue},
  urlcolor={Blue},
  pdfcreator={LaTeX via pandoc}}

\newcommand{\runninghead}{A Preprint }
\renewcommand{\runninghead}{A submission to the VIEWS 2023/24 Prediction
Competition }
\title{Forecasting Densities of Fatalities from State-based Conflicts
using Observed Markov Models\thanks{The research was funded by the
European Research Council, project H2020-ERC-2015-AdG 694640 (ViEWS) and
Riksbankens Jubileumsfond, grant M21-0002 (Societies at Risk)}}
\def\asep{\\\\\\ } % default: all authors on same column
\author{\textbf{David
Randahl}~\orcidlink{0000-0003-1069-6067}\\\\Department of Peace and
Conflict Research, Uppsala
University\\\\\href{mailto:david.randahl@pcr.uu.se}{david.randahl@pcr.uu.se}\asep\textbf{Johan
Vegelius}~\orcidlink{0000-0002-0497-0659}\\\\Department of Medical
Science, Uppsala University\\\\}
\date{}
\begin{document}
\maketitle
\begin{abstract}
In this contribution to the VIEWS 2023 prediction challenge, we propose
using an observed Markov model for making predictions of densities of
fatalities from armed conflicts. The observed Markov model can be
conceptualized as a two-stage model. The first stage involves a standard
Markov model, where the latent states are pre-defined based on domain
knowledge about conflict states. The second stage is a set of regression
models conditional on the latent Markov-states which predict the number
of fatalities. In the VIEWS 2023/24 prediction competition, we use a
random forest classifier for modeling the transitions between the latent
Markov states and a quantile regression forest to model the fatalities
conditional on the latent states. For the predictions, we dynamically
simulate latent state paths and randomly draw fatalities for each
country-month from the conditional distribution of fatalities given the
latent states. Interim evaluation of out-of-sample performance indicates
that the observed Markov model produces well-calibrated forecasts which
outperform the benchmark models and are among the top performing models
across the evaluation metrics.
\end{abstract}
{\bfseries \emph Keywords}
\def\sep{\textbullet\ }
Markov models \sep Forecasting \sep Density forecasts \sep 
Fatalities

\begin{figure}

\begin{minipage}{0.50\linewidth}
\includegraphics{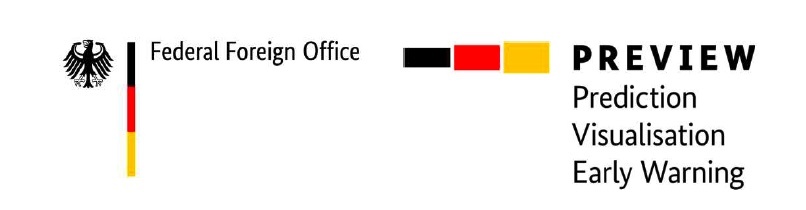}\end{minipage}%
\begin{minipage}{0.50\linewidth}
\includegraphics[width=0.5\textwidth,height=\textheight]{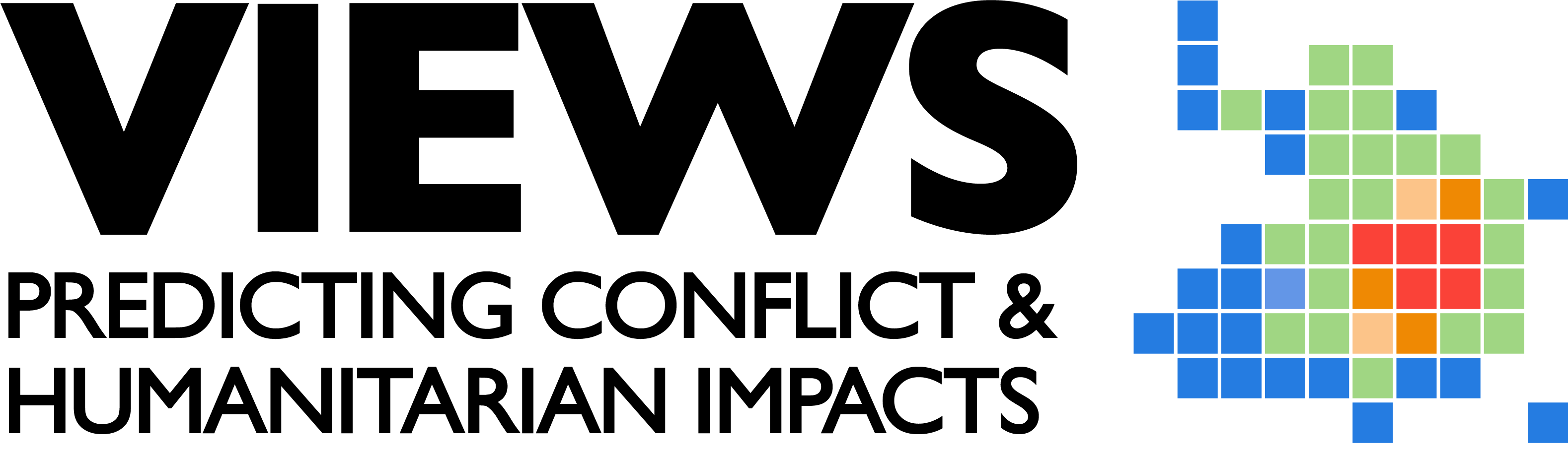}\end{minipage}%

\end{figure}%

\newpage

\section{Introduction}\label{sec-intro}

This paper describes a contribution to the VIEWS 2023/24 prediction
challenge.\footnote{Financial support for the Prediction Challenge was
  provided by the German Ministry for Foreign Affairs. For more
  information on the Prediction Challenge please see Hegre et. al
  (forthcoming) and
  https://viewsforecasting.org/research/prediction-challenge-2023} The
goal of the prediction challenge is for different teams to provide true
out-of-sample forecasts of the number of fatalities from state-based
conflicts for each country-month from July 2024 to June 2025 \emph{with
uncertainty}, i.e.~a full predictive density for each country month
(Hegre Håvard and et.al forthcoming). In this contribution we propose
using observed Markov models to forecast these densities.

Markov models have been widely applied across fields such as finance,
biology, medicine, and conflict research (see for instance Mor, Garhwal,
and Kumar 2021; Jackson 2011; Schrodt 2006). They have also been used in
previous VIEWS prediction competitions to forecast changes in fatality
levels from armed conflicts in Africa (Randahl and Vegelius 2022). In
this paper, we extend the observed Markov model introduced in the
previous competition to make distributional forecasts of fatalities.
Evaluation of the model's performance during the interim period shows
that the observed Markov model produced well-calibrated forecasts and
that they outperform all benchmark models across the evaluation metrics
and that they are among the top performing models overall.

Markov models provide an intuitive framework for conflict forecasting,
as different conflict dynamics can be viewed as different latent
conflict states. For instance, a country can be in a `peace' state, a
`low-intensity conflict' state, or a `war' state. As a country moves
through time, it can transition between these states but these
transitions will be conditional on current state, not on previous
states. This means that a country in the `peace' state is more likely to
remain in peace than a country in the `war' state, even if all other
factors are the same. In a standard Markov model, these states are
directly observed and transitions between explicitly modeled, while in a
hidden Markov models infer latent states from the data. The observed
Markov model differs from a standard Markov model in that the states of
the model are not necessarily of interest, instead the states are used
to condition the prediction of the outcome of interest. The models also
differ from \emph{hidden} Markov models in that the states are
researcher-defined and observed rather than inferred from the data.

We argue that the real-world dynamics of armed conflict closely resemble
a Markov model with a large number of states through which the countries
can transition over time. While these conflict-states in the real world
are not directly observable, we believe that they can be usefully
approximated for the purpose of conflict forecasting through the number
of observed fatalities conflicts within the country. Imposing this
structure can also be seen as a form of regularization, as it reduces
the complexity of the problem by reducing the number of possible states
the model has to consider and may thus lead to more stable and accurate
forecasts. Previous work has found that the observed Markov model
outperformed the hidden Markov model in forecasting changes in the
number of fatalities from state-based conflicts in Africa (Randahl and
Vegelius 2022).

In the following sections, we describe the VIEWS 2023/24 prediction
challenge and the fundamental idea of the observed Markov model,
followed by details on how we implement the model in practice to
forecast densities of fatalities from state-based conflicts. Lastly, we
present the results from the interim evaluation of the model and discuss
the implications of the results.

\section{The VIEWS 2023/24 Prediction
Challenge}\label{sec-prediction-challenge}

The VIEWS 2023/24 prediction challenge is a successor to the VIEWS
prediction competition (Vesco et al. 2022; H. Hegre, Vesco, and Colaresi
2022). Unlike the previous VIEWS prediction competition, focused on
predicting changes in the number of battle-related deaths, the VIEWS
2023/24 prediction challenge's aim is to predict the number of
battle-related deaths in each country in each month \emph{with
uncertainty} (Hegre Håvard and et.al forthcoming).\footnote{The teams
  are allowed to choose between two levels of analysis for their
  forecasts, either the country-month level or the prio-grid month. This
  contribution provides forecasts on the country-month level.} In the
prediction challenge, participating teams are asked to make true
out-of-sample predictions of the number of battle-related fatalities
from state-based conflicts in each country and month from July 2024 to
June 2025. The predictions were made in the form random draws from the
predicted density of the outcome of interest. The forecasts were
submitted to the VIEWS team by the end of June 2024 and are securely
stored until the actual outcomes are known, thus ensuring that the
forecasts are truly out-of-sample. In addition to the true out-of-sample
forecasts, participants are also asked to provide out-of-sample
predictive densities for an interim evaluation period in 2018-2023. The
interim predictions were submitted calendar year by calendar year using
only the data that was available up until October of the previous year.
This mimics the real-time forecasting challenge where we were asked to
produce forecasts for yet unknown outcomes. The VIEWS 2024/25
forecasting challenge is a \emph{live} forecasting challenge, meaning
that the forecasts are evaluated in real-time as the actual outcomes
become known.

The submitted models are evaluate across a range of different metrics
which all take the density of the forecasts into account. Specifically,
the predictive density is evaluated using the continuous ranked
probability score (CRPS), the ignorance score\footnote{Also known as the
  \emph{log-score}} (IGN), and the Mean Interval Score (MIS) (Hegre
Håvard and et.al forthcoming).\footnote{For a detailed description of
  the evaluation metrics, please see Hegre et. al. (forthcoming).} The
ranking of the models on the true out-of-sample forecasts can be
followed on the \href{https://predcomp.viewsforecasting.org/}{VIEWS
website}.

\subsection{Predictor features}\label{predictor-features}

As part of the prediction challenge, the VIEWS team provides a set of
predictor features from the main VIEWS data which the teams can use to
make their forecasts (H. Hegre et al. 2021, 2019). Specifically, a set
of 123 features were made available consisting of a mix of conflict
history, political, economic, demographic, and environmental features.
The features are provided at the country-month level and are available
from January 1990 to June 2024. The teams are allowed to use any of the
features in the prediction challenge and are also allowed to use any
external data sources they see fit.

\subsection{Benchmark models}\label{benchmark-models}

In addition to the predictor features, the VIEWS team also provides a
set of benchmark models which performance the teams can use as point of
reference for their own models. For the country-month level, for
different benchmark models are provided: two simple models where one
predicts zero fatalities across all country-months, and one which simply
predicts a poisson distribution with \(\lambda\) equal to the number of
fatalities in the previous month; and two more complex models which use
the distribution of fatalities in the last 12 months and 240 months
respectively.\footnote{For a detailed description of the benchmark
  models, please see Hegre et. al. (forthcoming)}

\section{Observed Markov Models}\label{sec-observed-markov}

The observed Markov model is an extension of the standard Markov model
in which the latent Markov states are used to condition the modeling or
prediction of a real-valued outcome of interest beyond the the states or
transition probabilities between them. In essence, the observed Markov
model can be seen as a two-stage model in which the first stage is a
standard Markov model with a set of pre-defined latent states and the
second stage is a set of regression models conditional on the latent
states.

Standard Markov models are used to represent Markov processes, which
involve a set of states through which units move over time. At any given
time, each unit occupies one of these states, and the unit can either
remain in the same state or transition to another according to some
probabilities. The sequence of state transitions forms what is known as
a Markov chain. A key assumption of the Markov process is the Markov
property, which stipulates that the probability of transitioning to the
next state depends only on the current state, not on the prior sequence
of states. In this framework, states are directly observed, and
transition probabilities can be estimated simply by counting the
observed transitions between states in the data. The resulting
transition matrix, derived from these counts, can be used to predict
future states by multiplying the current state vector by the transition
matrix, repeating this process for as many time steps as needed
(Fosler-Lussier 1998; Jackson 2011).

While transition probabilities can be calculated based on observed
transitions alone, it is also possible to estimate the transition matrix
using additional covariates. In this case, the transitions between
states can be viewed as a simple classification problem where the
probabilities of future states are modeled as conditional on the current
state \emph{and} some covariates. These transition probabilities are
commonly estimated using standard parametric models such as logistic
regression models (Jackson 2011) but can be estimated using any
classification model, including non-parametric and machine-learning
models (Randahl and Vegelius 2022).

One limitation of the standard Markov model is that it only forecasts
states and transition probabilities, making it unsuitable for predicting
real-valued outcomes. An alternative is to use a \emph{hidden} Markov
model. In the hidden Markov model, the states of the process are not
directly observed but are instead inferred from some other observable
outcome. In the hidden Markov model, each of the latent states are
associated with a parametric distribution of the observed outcome, which
may also be conditioned on some covariates, and the state and transition
probabilities between the states are estimated using the
forward-backward algorithm. This allows the hidden Markov model to be
used to forecast the latent states of the process as well as the
real-valued outcome of interest (Fosler-Lussier 1998; Jackson 2011).
However, hidden Markov models are more complex and require large amounts
of data to accurately estimate the states and transitions. Additionally,
they are constrained by parametric assumptions that may not capture the
true distribution of the observed outcome.

The observed Markov model aims to combine the strengths of the standard
Markov model and the hidden Markov model by applying the logic of hidden
Markov models on processes with observed or imposed states. In this
model, the states of the process are not the main focus of the model but
are instead used to condition the prediction of a real-valued outcome.
The states can either be directly observable or defined by the
researcher based on domain knowledge and artificially imposed.
Transition probabilities between states can be estimated using any
classifier, and the real-valued outcome is modeled separately for each
state, using any type of regression model, including non-parametric and
machine-learning models. Unlike the hidden Markov model, the observed
Markov model is not restricted by parametric assumptions, allowing for
more flexibility in forecasting both states and outcomes, while
remaining relatively simple to estimate and interpret. This allows the
observed Markov model to be used to forecast both the latent states of
the process and the real-valued outcome, while remaining relatively
simple to estimate and interpret. In essence, this makes the observed
Markov model a two-stage model in which the first stage is a
classification model that predicts the states of the process and the
second stage is a regression model that predicts the real-valued outcome
conditional on the states.

We believe the observed Markov model is particularly useful for
prediction problems where strong domain knowledge exists, and where the
data is too sparse for a hidden Markov model to be effective. In such
cases, defining states based on domain knowledge can act as a form of
regularization, leading to more stable and accurate predictions compared
to the hidden Markov model, which needs to learn the states and
transition probabilities from the data. The observed Markov model is
also more flexible than the hidden Markov model because it does not
require explicit parametric assumptions about state distributions and
allows different models and covariates for each state and transition.
This makes it a powerful tool for scenarios where the outcome of
interest depends on the states of the process and where states can be
informed by domain knowledge.

\section{Forecasting densities of fatalities using the observed Markov
model}\label{forecasting-densities-of-fatalities-using-the-observed-markov-model}

To make predict an outcome of interest \emph{with uncertainty}, i.e.~to
produce a predictive density, we suggest pairing the observed Markov
model with a dynamic simulation approach. In this approach, the Markov
chain is simulated by randomly drawing future states based on the
transition probabilities, conditional on the current state and its
associated model and feature values. The real-valued outcome of interest
is then predicted by conditioning on these simulated states and their
corresponding features. This process is repeated for a large number of
simulations, generating a range of possible outcomes for each time
period.

If the goal is point predictions, the mean or median of the simulated
values can be used. However, when predicting with uncertainty, point
predictions alone are insufficient. Instead, the distribution of the
outcome must be considered. One option is to assume a parametric
distribution around the point prediction in each state, using this to
forecast outcomes across the simulations. Alternatively, regression
models that generate explicit densities can be employed, and these
densities can be combined to produce a predictive density. Lastly, a
quantile regression model can be used to draw random quantiles from the
distribution of the outcome in each time period. This method creates a
broad range of potential values that can be used to construct a
predictive density.

In this contribution, we have chosen to use the quantile regression
forest to produce predictive densities for the outcome of interest in
each time period as this approach is straightforward to implement in the
context of the observed Markov model and works for a wide range of
regression models.

\subsection{Markov states}\label{markov-states}

The first step in the observed Markov model is to define the states of
the process. In this contribution, we have chosen to define four
distinct states based on the number of fatalities in the current and
previous time period.\footnote{This follows the approach taken by
  Randahl and Vegelius (2022) in the previous VIEWS prediction
  competition} The states are defined as follows:

\textbf{Peaceful}: No fatalities in the current or previous time period.

\textbf{Escalation}: No fatalities in the previous time period but
fatalities in the current time period.

\textbf{War}: Fatalities in the previous time period and fatalities in
the current time period.

\textbf{De-escalation}: Fatalities in the previous time period but no
fatalities in the current time period.

While these states are a simplification of the true dynamics of the
process, they are intended to capture the general patterns of escalation
and de-escalation in the number of fatalities in the process. In
particular, letting the states \textbf{Escalation} and
\textbf{De-escalation} to be transient states which countries can only
stay in for one time-period is expected provide much more stable
predictions in the two non-transient states \textbf{Peaceful} and
\textbf{War} as countries cannot transition directly between these two
states. An additional upside with defining the \textbf{Escalation} and
\textbf{De-escalation} states as transient is that given a specific
state in the current time-period, only two possible states can be
reached in the next time-period. This simplifies the estimation as the
transitions between states is a simple binary classification problem.

Defining the \textbf{Peaceful} and \textbf{De-escalation} states as
having no fatalities in the current time period also simplifies the
estimation of the real-valued outcome of interest as the outcome of
interest is defined to be zero in these states. This allows us to focus
on modeling the outcome of interest in the \textbf{Escalation} and
\textbf{War} states where the outcome of interest is defined to be
non-zero.

\subsection{Transition and outcome
models}\label{transition-and-outcome-models}

Both the transition probabilities and the outcome of interest are
modeled using random forest models. For the transitions between states,
we use a standard random forest classifier with probability estimates,
and for the outcome of interest, we use a quantile regression forest
(Breiman 2001; Meinshausen and Ridgeway 2006). The random forest models
are chosen for their flexibility and ability to model complex
relationships in the data without making strong parametric assumptions.

As input features to the models, we use a set of principal components
(PCs) extracted by thematically grouping the features in the
data.\footnote{Details on the features included in the data can be found
  in (Hegre Håvard and et.al forthcoming)} Specifically, we use two PCs
for the political features from the Varieties of Democracy (VDEM)
dataset, two PCs for the violence history features, two PCs for the
development features from the World Development Indicators (WDI), one PC
for the military expenditure, one PC for demographics, one PC for
environmental features, and two PCs for the neighborhood conflict
features. Additionally, we include a decay function for the total number
of fatalities earlier in the unit's history with a half-life of 12
months. This decay function is included to capture the long-term
dynamics of the process and to allow for the model to learn from the
history of the process. We use the same set of features for both the
transition and outcome models.

\section{Evaluation}\label{sec-evaluation}

The performance of the models in the interim evaluation period,
2018-2023, from all teams participating in the prediction
competition\footnote{Details on the competing models are available
  (Hegre Håvard and et.al forthcoming)} in is shown in Table
\ref(tab:results) below. The results show that the OMM model does well
in comparison to the other models in the competition, with the
second-best CRPS score and the best IGN score. On the MIS score, the OMM
performs slightly worse, but still better than the majority of the other
models.

\begin{table}[H]
\begin{tabular}{lllll}
\label{tab:results}
\textbf{Team}                     & Model                          & crps       & ign        & mis        \\
\hline
Pace                              & ShapeFinder                    & 44,86  & 0,68 & 806,67 \\
\textbf{Markov} & \textbf{OMM} & \textbf{45,9} & \textbf{0,58} & \textbf{868,26} \\
ConflictForecast                   & conflictforecast\_v2           & 47,54 & 0,62 & 769,84 \\
Temporal Fusion Transformers CCEW & TFT                            & 47,90  & 0,74 & 836,07 \\
Muchlinski and Thornhill          & views\_zip                     & 48,26 & 1,71 & 936,59  \\
UNITO                             & unito\_transformer             & 48,90 & 0,96 & 861,33 \\
benchmark                         & benchmark\_conflictology\_12m  & 49,36 & 0,65 & 873,53 \\
markov                            & hpmm                           & 49,50 & 0,70 & 846,88 \\
kit\_becker\_drauz                & quantile\_forecast             & 54,52 & 0,65 & 942,92 \\
markov                            & gpcmm                          & 55,01   & 8,90 & 975,74 \\
bodentien\_rueter                 & bodentien\_rueter\_negbin      & 56,11 & 0,62 & 899,04 \\
benchmark                         & benchmark\_boot\_240           & 56,17 & 1,14 & 1101,02 \\
benchmark                         & benchmark\_exactly\_zero       & 56,84 & 1,59 & 1136,80 \\
Brandt-UTDallas                   & P\_GLMM                        & 111,60 & 1,10 & 2210,64 \\
Brandt-UTDallas                   & Neg\_Bin\_GLMM                 & 132,05 & 1,08 & 2277,12 \\
Brandt-UTDallas                   & TW\_GLMM                       & 141,73 & 0,92 & 2743,37 \\
benchmark                         & benchmark\_last\_with\_poisson & 158,16 & 1,14  & 3137,09 \\
Brandt-UTDallas                   & TW\_GAM                        & 1585645,09 & 1,06 & 31320347,3 \\
Brandt-UTDallas                   & Neg\_Bin\_GAM                  & 9460460,1  & 0,72 & 186598021  \\
Brandt-UTDallas                   & P\_GAM                         & 44311833,1 & 2,02  & 886236557 
\end{tabular}
\end{table}

Overall, the results show that the OMM model is competitive with the
other models in the competition, and it performs very well on two of the
three evaluation metrics. Importantly, the OMM is one of only two
models, together with the ConflictForecast model, which outperforms all
four benchmark models across all three evaluation metrics.

\section{Discussion}\label{sec-discussion}

This research note has presented the Observed Markov Model for
predicting fatalities from armed conflict with uncertainty in the VIEWS
2023/24 prediction competition. The OMM model is a novel approach to
conflict forecasting that combines a Markov model for the transitions
between states with a random forest model for the outcome of interest.
The model is designed to capture the complex dynamics of armed conflict
and to provide accurate and reliable predictions of the onset of armed
conflict. The results from the interim evaluation period show that the
OMM model performs well in comparison to the other models in the
competition, with the second-best CRPS score and the best IGN score.
These results suggest that the OMM model is a promising approach for
forecasting fatalities from armed conflict, and it is likely to be a
valuable tool for policymakers and researchers in the field of conflict
forecasting.

This submission should, however, only be considered a first step in the
use of the OMM model for conflict forecasting. There are several areas
where the model could be further developed and improved. For example,
the model presented in this research note only use a small number of
principle components of features known to be linked to armed conflict as
input to the model. Future work could improve the model by including
additional features and data sources to capture more of the complexity
of armed conflict, or using theory to guide the selection of features.
Different features could also be selected for the transition and
regression models to improve performance further. Additionally, the OMM
model presented here uses only four Markov states which are defined
based on the number of fatalities in the current and previous month.
Further work could theorize and test different number of states and
state definitions to improve the model's performance.

\newpage{}

\section*{References}\label{references}
\addcontentsline{toc}{section}{References}

\phantomsection\label{refs}

\end{document}